\documentclass[aps,print,twocolumn,showpacs,superscriptaddress]{revtex4}%
\usepackage{amsfonts}
\usepackage{amsmath}
\usepackage{amssymb}
\usepackage{graphicx}%
\providecommand{\U}[1]{\protect\rule{.1in}{.1in}}

\begin{document}
\title{Spin current and polarization reversal through a single-molecule magnet with
ferromagnetic electrodes}
\author{Haiqing Xie}
\affiliation{Institute of Theoretical Physics and Department of Physics, Shanxi University,
Taiyuan 030006, China}
\author{Qiang Wang}
\affiliation{Institute of Theoretical Physics and Department of Physics, Shanxi University,
Taiyuan 030006, China}
\author{Bo Chang}
\affiliation{College of Physics and Optoelectronics, Taiyuan University of Technology,
Taiyuan 030024, China}
\author{Hujun Jiao}
\affiliation{Institute of Theoretical Physics and Department of Physics, Shanxi University,
Taiyuan 030006, China}
\author{J.-Q. Liang}
\email{jqliang@sxu.edu.cn}
\affiliation{Institute of Theoretical Physics and Department of Physics, Shanxi University,
Taiyuan 030006, China}

\begin{abstract}
We theoretically study the spin-polarized transport through a single-molecule
magnet, which is weakly coupled to ferromagnetic leads, by means of the
rate-equation approach. We consider both the ferromagnetic and
antiferromagnetic exchange-couplings between the molecular magnet and
transported electron-spin in the nonlinear tunneling regime. For the
ferromagnetic exchange-coupling, spin current exhibits step- and basin-like
behaviors in the parallel and antiparallel configurations respectively. An
interesting observation is that the polarization reversal of spin-current can
be realized and manipulated by the variation of bias voltage in the case of
antiferromagnetic exchange-coupling with antiparallel lead-configuration,
which may be useful in the development of spintronic devices, while the bias
voltage can only affect the magnitude of spin-polarization in the
ferromagnetic coupling.

\end{abstract}
\date{\today}
\maketitle

\section{Introduction}

In the past few years, electron transport through magnetic molecules,
especially the single-molecule magnet (SMM), was intensively studied in both
experimental\cite{exp1,exp2,exp3,exp4,Timm0} and
theoretical\cite{Kim,Timm1,Timm2,Timm3,Timm4,Kondo,JB11,JB12,JB13,Shen,JB2,JB3,JB4,
Leuenberger,Martin1,Martin2,RWang,Xing} aspects, which is stimulated by the
fundamental importance as well as potential applications in molecular
spintronics. Many fascinating properties have been found, such as complex
tunneling spectra\cite{Kim,Timm1}, negative differential conductance
(NDC)\cite{exp1,Timm2,Timm3,JB2}, Kondo effect\cite{Kondo,JB4}, Berry phase
blockade\cite{Leuenberger}, full counting statistics\cite{Martin1}, colossal
spin fluctuations\cite{Martin2}, and so on, which are resulted from the large
spin-number and high anisotropy of SMM. The complete current suppression and
negative differential conductance are also confirmed by the
single-molecule-transistor measurement, which as a function of bias, gate
voltage, and external magnetic field provide evidences of magnetic signatures
of the SMM\cite{exp1}. Manipulation of spins in magnetic molecules, which is
based on the development of spin-controlling techniques, may result in new
strategies of quantum-state control. It is demonstrated that the charge and
spin states of molecule can be identified from the measured tunneling spectra
due to the existence of exchange interaction between the\textbf{ }local spin
of magnetic molecule and the spin of tunneling electron, particularly the
antiferromagnetic exchange interaction in a magnetic single-molecule
transistor based on N@C60\cite{exp3,Timm0}. On the other hand the tunneling
current between magnetic electrodes can also control the orientation of
molecular magnet shown by the study of colossal spin
fluctuations\cite{Martin2}.

\begin{figure}[ptb]
\centering \vspace{0cm} \hspace{0cm}
\scalebox{0.85}{\includegraphics{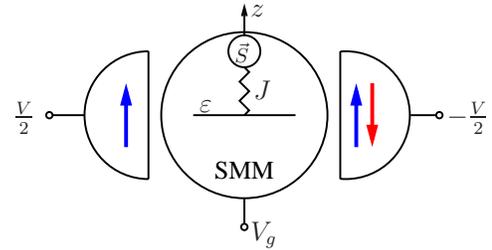}}\caption{(Color online) The
schematic diagram of quantum transport through a SMM.}%
\label{fig1}%
\end{figure}

Based on the study of spin polarized transport through a SMM some new
spintronic devices are proposed, which display many interesting effects. In
particular the spin-diode behavior\cite{JB2} and tunnel magnetoresistance
(TMR) \cite{JB3,JB4} are shown to be resulted from the exchange coupling
between the lowest-unoccupied-malecular-orbital (LUMO) level and the core
spin. Furthermore a large negative TMR is predicted in the case of
antiferromagnetic exchange-coupling \cite{JB3}. The giant spin amplification
and spin-blockade behavior are useful in molecular spintronics to serve as a
read-out mechanism \cite{Timm2,Timm3}. It is also found that the magnetization
of SMM can be reversed by spin polarized current \cite{Timm2,JB11,JB12,JB13}
and spin-bias driven magnetization reversal is also observed
theoretically\cite{Shen}. The most recent study is focused on the manipulation
of SMM by thermal spin-transfer torque \cite{Xing}. Moreover a highly
spin-polarized current can be generated by thermoelectric effects\cite{RWang}.

To date, most researches are concentrated on the manipulation of a SMM by the
transported electron, which on the other hand may be used as a probe to
explore the level structure of the SMM. The spin-polarized current through a
SMM itself has received little attention, while in the quantum-dot (QD) case
it is extensively studied\cite{Konig,Jauho1,Jauho2} in terms of nonequilibrium
Green's functions\cite{Jauho1} and master-equation \cite{Jauho2} respectively.
The QD acts as a spin-current diode giving rise to the spin blockade. The spin
current in QD system shows quite different characteristics compared with the
charge transport \cite{GangSu,Jauho1,Yuan}.

We in the present paper study the spin-polarized transport through a SMM,
which has more complex level-structure than the single QD. Moreover the
exchange coupling between SMM and the transported electron-spin leads to
additional dynamic mechanism to manipulate the spin-polarization of current.
Both parallel and antiparallel configurations of lead magnetization are
considered along with the ferromagnetic and antiferromagnetic couplings. For
the ferromagnetic exchange-coupling, the spin current variation with respect
to the bias voltage exhibits a step-like curve in parallel configurations and
basin-like behavior in the antiparallel configurations, respectively. The
interesting observation in antiferromagnetic coupling case is that, the
spin-polarization can be reversed with increasing bias voltage in antiparallel lead-configurations.

\section{Exchange coupling and transition rate of tunneling}

We consider a system which consists of a SMM coupled to two ferromagnetic
metallic-electrodes (see Fig. 1.), which can be described by the Hamiltonian
\cite{Timm2,Timm3,JB12}%
\begin{equation}
H=H_{SMM}+H_{leads}+H_{T}.
\end{equation}
The first term in Eq. (1) concerning the SMM of easy-axis anisotropy with
parameter $K_{z}>0$ has the form
\begin{equation}
H_{SMM}=\sum_{\sigma}(\varepsilon-eV_{g})d_{\sigma}^{\dag}d_{\sigma
}+Ud_{\uparrow}^{\dag}d_{\uparrow}d_{\downarrow}^{\dag}d_{\downarrow
}-J\mathbf{s}\cdot\mathbf{S}-K_{z}(S^{z})^{2}\text{,}%
\end{equation}
in which $J\mathbf{s}\cdot\mathbf{S}$\ is the exchange interaction between
electron spin and the giant spin $\mathbf{S}$ of SMM with $J$ being the
exchange coupling parameter and $\mathbf{s}\mathbf{\equiv}$ $\sum
_{\sigma\sigma^{\prime}}d_{\sigma}^{\dag}(\mathbf{\sigma}_{\sigma
\sigma^{\prime}}/2)d_{\sigma^{\prime}}$ is the corresponding spin operator of
electron ($\sigma$ is the vector of Pauli matrices). $d_{\sigma}^{\dag}%
$($d_{\sigma}$) denotes the relevant electron creation (annihilation) operator
and $\varepsilon$ is the single-electron energy of the LUMO level, which is
tunable by the gate voltage $V_{g}$. $U$ represents the Coulomb interaction of
two electrons of opposite spins. The exchange interaction can be of either
ferromagnetic ($J>0$) or antiferromagnetic ($J<0$) type.

\begin{figure}[ptb]
\centering \vspace{0cm} \hspace{0cm}
\scalebox{0.7}{\includegraphics{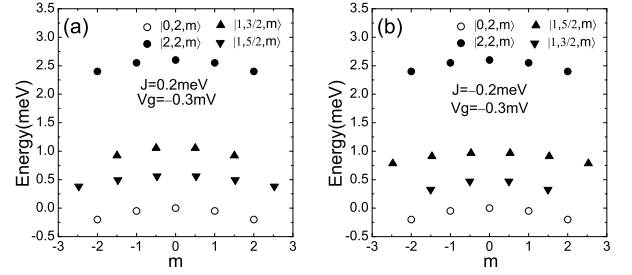}}\caption{(Color online) Energy
spectra of the SMM for ferromagnetic (a) and antiferromagnetic (b) exchange-
interactions as a function of magnetic quantum-number $m$ with $S=2$,
$K_{z}=0.05$meV, $U=1$meV, $Vg=-0.3$mV.}%
\label{fig2}%
\end{figure}

The many-body states of electron-spin and molecule are expressed in terms of
the eigenstates of operator $\mathbf{S}_{tot}^{z}$, $\left\vert n,S_{tot}%
;m\right\rangle $, where $\mathbf{S}_{tot}$=$\mathbf{s+S}$ denotes the total
spin operator of electron and molecule, with $n$ denoting the charge state of
the SMM, $S_{tot}$ the total spin quantum-number and $m$ the eigenvalues of
$\mathbf{S}_{tot}^{z}$\cite{Timm2,JB12,JB3,Martin2}. For the case of
$n=0$,\ the state $\left\vert 0,S_{tot}=S;m\right\rangle \equiv\left\vert
0\right\rangle _{orb}\otimes\left\vert m\right\rangle _{mol}$ and the
corresponding eigenenergy is $\varepsilon_{\left\vert 0,S_{tot}%
=S;m\right\rangle }=-K_{z}m^{2}$, while when $n=2$, the state is $\left\vert
2,S_{tot}=S;m\right\rangle \equiv\left\vert \uparrow\downarrow\right\rangle
_{orb}\otimes\left\vert m\right\rangle _{mol}$ and the eigenenergy is
$\varepsilon_{\left\vert 2,S_{tot}=S;m\right\rangle }=2(\varepsilon
-eV_{g})+U-K_{z}m^{2}$. For $n=1$ and $K_{z}-J/2>0$, the state and eigenenergy
are found as $\left\vert 1,S_{tot}=S\pm1/2;m\right\rangle \equiv
a_{m\downarrow}^{\left(  \pm\right)  }\left\vert \downarrow\right\rangle
_{orb}\otimes\left\vert m+1/2\right\rangle _{mol}+b_{m\uparrow}^{\left(
\pm\right)  }\left\vert \uparrow\right\rangle _{orb}\otimes\left\vert
m-1/2\right\rangle _{mol}$ and $\varepsilon_{\left\vert 1,S_{tot}%
=S\pm1/2;m\right\rangle }=(\varepsilon-eV_{g})+J/4-K_{z}(m^{2}+1/4)\pm\Delta
E(m)$ respectively. For $n=1$ and $K_{z}-J/2<0$, we have $\left\vert
1,S_{tot}=S\mp1/2;m\right\rangle \equiv a_{m\downarrow}^{\left(  \pm\right)
}\left\vert \downarrow\right\rangle _{orb}\otimes\left\vert m+1/2\right\rangle
_{mol}+b_{m\uparrow}^{\left(  \pm\right)  }\left\vert \uparrow\right\rangle
_{orb}\otimes\left\vert m-1/2\right\rangle _{mol}$, and $\varepsilon
_{\left\vert 1,S_{tot}=S\mp1/2;m\right\rangle }=(\varepsilon-eV_{g}%
)+J/4-K_{z}(m^{2}+1/4)\pm\Delta E(m)$, where $\Delta E(m)=[K_{z}(K_{z}%
-J)m^{2}+(J/4)^{2}(2S+1)^{2}]^{1/2}$, and $a_{m\downarrow}^{\left(
\pm\right)  }$, $b_{m\uparrow}^{\left(  \pm\right)  }$ are effective
Clebsch--Gordan coefficients\cite{JB12}.

The SMM is weakly coupled to two ferromagnetic metallic leads with the
Hamiltonian given by
\[
H_{leads}=\sum_{\mathbf{\alpha=L,R}}\sum_{\mathbf{k}\sigma}\varepsilon
_{\alpha\mathbf{k}}c_{\alpha\mathbf{k}\sigma}^{\dag}c_{\alpha\mathbf{k}\sigma
},
\]
where $c_{\alpha\mathbf{k}\sigma}^{\dag}$ ($c_{\alpha\mathbf{k}\sigma}$) is
the creation (annihilation) operator for an electron of spin-index $\sigma$
and wave vector $\mathbf{k}$ in the $\alpha$ lead. The spin polarization of
ferromagnetic lead $\alpha$ is defined as $P_{\alpha}=(\rho_{\alpha+}%
-\rho_{\alpha-})/(\rho_{\alpha+}+\rho_{\alpha-})$, with $\rho_{\alpha+(-)}$
denoting the density of states for the majority (minority) electrons in the
$\alpha$ lead. In this paper, we assume that the magnetization of
ferromagnetic leads is collinear with the easy axis of SMM. The tunneling
processes between the molecule and leads can be described by Hamiltonian
\[
H_{T}=\sum_{\alpha\mathbf{k}\sigma}[t_{\alpha}c_{\alpha\mathbf{k}\sigma}%
^{\dag}d_{\sigma}+t_{\alpha}^{\ast}d_{\sigma}^{\dag}c_{\alpha\mathbf{k}\sigma
}],
\]
with $t_{\alpha}$ denoting the tunnel matrix element between the molecule and
$\alpha$ lead. The spin-dependent tunnel coupling-strength is given by
\[
\Gamma_{\alpha\sigma}=2\pi\rho_{\alpha\sigma}\left\vert t_{\alpha}\right\vert
^{2},
\]
and the total coupling-strength is $\Gamma_{\alpha}=\sum_{\sigma}%
\Gamma_{\alpha\sigma}$. In order to obtain transport properties in both the
sequential and cotunneling regimes, we employ the rate-equation approach with
the help of $T$-matrix \cite{Timm5,Koch1,Koch2,Bruus}, which satisfies the
iterative equation
\begin{equation}
T=H_{T}+H_{T}\frac{1}{E_{I}-H_{0}+i0^{+}}T,
\end{equation}
where $E_{I}$ is the energy of initial state $\left\vert I\right\rangle $, and
$0^{+}$ denotes a small quantity in the retarded Green function. The
transition strength from initial state $\left\vert I\right\rangle $ to final
state $\left\vert F\right\rangle $ can be evaluated by the perturbation
expansion of $T$-matrix in terms of the generalized Fermi's golden
rule\cite{Bruus},
\begin{align}
\Gamma_{FI} &  =\frac{2\pi}{\hbar}\left\vert \left\langle F\left\vert
T\right\vert I\right\rangle \right\vert ^{2}\delta(E_{F}-E_{I})\nonumber\\
&  =\frac{2\pi}{\hbar}\left\vert \left\langle F\right\vert H_{T}+H_{T}\frac
{1}{E_{I}-H_{0}+i0^{+}}H_{T}+...\left\vert I\right\rangle \right\vert
^{2}\nonumber\\
&  \delta(E_{F}-E_{I}),
\end{align}
where $\left\vert I\right\rangle $ and $\left\vert F\right\rangle $ are
actually product states of the electron-lead and molecule. After eliminating
the lead degree of freedom, the sequential transition rate up to the
second-order of tunneling Hamiltonian $H_{T}$ can be obtained from the
transition-strength formula%
\begin{align}
W_{\alpha\sigma}^{i,i^{\prime}} &  =\frac{\Gamma_{\alpha\sigma}}{\hbar
}[f_{\alpha}(\varepsilon_{i^{\prime}}-\varepsilon_{i}-\mu_{\alpha})\left\vert
\left\langle i^{\prime}\right\vert d_{\sigma}^{\dag}\left\vert i\right\rangle
\right\vert ^{2}\nonumber\\
&  +[1-f(\varepsilon_{i}-\varepsilon_{i^{\prime}}-\mu_{\alpha})]\left\vert
\left\langle i^{\prime}\right\vert d_{\sigma}\left\vert i\right\rangle
\right\vert ^{2}]\text{,}%
\end{align}
which describes the transition of molecule from state $\left\vert
i\right\rangle $ to $\left\vert i^{\prime}\right\rangle $, due to
spin-$\sigma$ electron tunneling into or out of lead $\alpha$. Where $\mu
_{L}=-eV/2$ , $\mu_{R}=eV/2$, $\varepsilon_{i}$ is the energy of state
$\left\vert i\right\rangle $ and $f(x)$ is the Fermi distribution function.
For the SMM model we can find the sequential transition selection-rule :
$\left\vert \Delta n\right\vert =1$, $\left\vert \Delta S_{tot}\right\vert
=1/2$, and $\left\vert \Delta m\right\vert =1/2$. The fourth-order cotunneling
transition rate%

\begin{align}
&  W_{\alpha\sigma,\alpha^{\prime}\sigma^{\prime}}^{i,i^{\prime}}\nonumber\\
&  =\frac{\Gamma_{\alpha\sigma}\Gamma_{\alpha^{\prime}\sigma^{\prime}}}%
{2\pi\hbar}\int d\varepsilon f(\varepsilon-\mu_{\alpha})[1-f(\varepsilon
-\varepsilon_{i^{\prime}}+\varepsilon_{i}-\mu_{\alpha^{\prime}})]\nonumber\\
&  \left\vert \sum_{j}\frac{\left\langle i^{\prime}\right\vert d_{\sigma
}^{\dag}\left\vert j\right\rangle \left\langle j\right\vert d_{\sigma^{\prime
}}\left\vert i\right\rangle }{\varepsilon-\varepsilon_{i^{\prime}}%
+\varepsilon_{j}-i0^{+}}+\frac{\left\langle i^{\prime}\right\vert
d_{\sigma^{\prime}}\left\vert j\right\rangle \left\langle j\right\vert
d_{\sigma}^{\dag}\left\vert i\right\rangle }{\varepsilon+\varepsilon
_{i}-\varepsilon_{j}+i0^{+}}\right\vert ^{2},
\end{align}
stands for the virtual transitions from molecular states $\left\vert
i\right\rangle $ to $\left\vert i^{\prime}\right\rangle $ while changing a
spin-$\sigma$ electron of lead $\alpha$ into spin-$\sigma^{\prime}$ electron
of lead $\alpha^{\prime}$.\textbf{ }The divergence of above expressions
existed whenever energy denominators vanish has to be eliminated by some
regularization procedures \cite{Koch1,Koch2,Matveev,Wegewijs,Timm6,Weymann}
and in this paper we adopt the method described in Ref .\cite{Weymann}. The
cotunneling transition selection rules are seen to be $\left\vert \Delta
n\right\vert =0$, and $\left\vert \Delta m\right\vert =0,\pm1$.

For the weak coupling between lead and molecule, i.e. $\Gamma_{\alpha\sigma
}\ll k_{B}T$, electron transport can be considered as a stochastic Markovian
process and thus the time evolution is described by the following rate equations%

\begin{align}
\frac{dP_{i}}{dt}  &  =\sum_{\alpha\alpha^{\prime}\sigma\sigma^{\prime
}i^{\prime}\neq i}-(W_{\alpha\sigma,\alpha^{\prime}\sigma^{\prime}%
}^{i,i^{\prime}}+W_{\alpha\sigma}^{i,i^{\prime}})P_{i}\nonumber\\
&  +\sum_{\alpha\alpha^{\prime}\sigma\sigma^{\prime}i^{\prime}\neq
i}(W_{\alpha\sigma}^{i^{\prime},i}+W_{\alpha^{\prime}\sigma^{\prime}%
,\alpha\sigma}^{i^{\prime},i})P_{i^{\prime}},
\end{align}
where $P_{i}$ is the probability of finding molecule in the many-body state
$i$. The stationary probabilities obtained from the condition, $\frac{dP_{i}%
}{dt}=0$, with the transition rate together result in the current of
spin-$\sigma$ through lead $\alpha$:
\begin{align}
I_{\alpha\sigma}  &  =-e(-1)^{\delta_{R\alpha}}\sum_{\alpha^{\prime}\neq
\alpha\sigma^{\prime}ii^{\prime}}[(n_{i^{\prime}}-n_{i})W_{\alpha\sigma
}^{i,i^{\prime}}P_{i}\nonumber\\
&  +(W_{\alpha\sigma,\alpha^{\prime}\sigma^{\prime}}^{i,i^{\prime}}%
-W_{\alpha^{\prime}\sigma^{\prime},\alpha\sigma}^{i,i^{\prime}})P_{i}],
\end{align}
from which we obtain the charge current $I_{\alpha}=I_{\alpha\uparrow
}+I_{\alpha\downarrow}$ and the spin current $I_{\alpha s}=I_{\alpha\uparrow
}-I_{\alpha\downarrow}$. The current polarization is defined by $\chi
=(I_{\alpha\uparrow}-I_{\alpha\downarrow})/(I_{\alpha\uparrow}+I_{\alpha
\downarrow})$.

\begin{figure}[ptb]
\centering \vspace{0cm} \hspace{0cm}
\scalebox{1.1}{\includegraphics{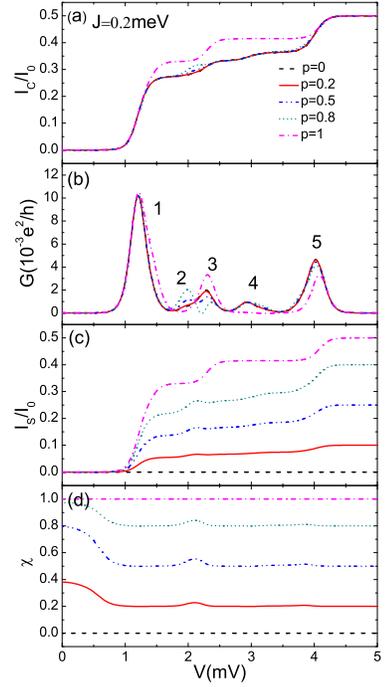}}\caption{(Color online) (a) Charge
current $I_{c}$, (b) differential conductance $G$, (c) spin current $I_{s}$,
and (d) current polarization $\chi$ as a function of the bias voltage $V$ for
different lead-polarizations $p$ and ferromagnetic exchange coupling
($J=0.2meV$) in case of parallel configuration of lead-magnetization.}%
\label{fig3}%
\end{figure}

\section{ Currents and polarizations in the nonlinear regime}

The numerical results of the charge current, differential conductance, spin
current, and current polarization in the nonlinear regime are carried out with
parameters chosen as $S=2$, $\varepsilon_{d}=0.5meV$, $Vg=-0.3mV$, $J=0.2meV$,
$U=1meV$, $K_{z}=0.05meV$ and $k_{B}T=0.04meV$. We set the tunnel coupling
between SMM and ferromagnetic leads to be $\Gamma=\Gamma_{L}=\Gamma
_{R}=0.001meV$ and assume the same polarizations of two leads i.e.,
$p_{L}=p_{R}=p$. In addition, the current and differential conductance are
scaled in units of $I_{0}=2e\Gamma/\hbar$ and $G_{0}=10^{-3}e^{2}/h$,
respectively. With the chosen parameters the numerical energy-spectrum of SMM
is obtained as in Fig.2, from which one can find that the energy-eigenvalue of
SMM-states corresponding to the total spin-number $S_{tot}=5/2$ is lower than
that of $S_{tot}=3/2$ for the ferromagnetic exchange-interaction (Fig. 2(a)),
while the situation is just opposite for the antiferromagnetic case (Fig.
2(b)). In what follows the spin currents are analyzed for two types of
exchange interactions with ferromagnetic leads of both parallel and
antiparallel configurations respectively.

\begin{figure}[ptb]
\centering \vspace{0cm} \hspace{0cm}
\scalebox{1.1}{\includegraphics{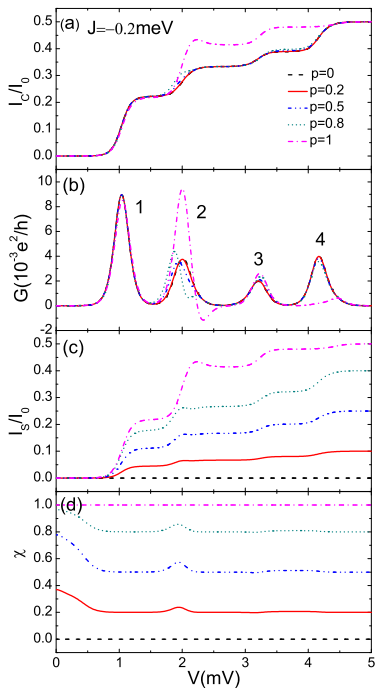}}\caption{(Color online) (a) Charge
current $I_{c}$, (b) differential conductance $G$, (c) spin current $I_{s}$,
and (d) current polarization $\chi$ as a function of the bias voltage $V$ for
different lead-polarizations $p$ and antiferromagnetic exchange coupling
($J=-0.2meV$) in case of parallel configuration of lead-magnetization.}%
\label{fig4}%
\end{figure}

\subsection{Parallel configuration}

For the ferromagnetic exchange-interaction, the bias-voltage dependences of
charge current $I_{c}$ and differential conductance $G$ for different
lead-polarizations $p$ are shown in Fig. 3(a) and Fig. 3(b) respectively,
where it is seen that the deviations of $I_{c}$ and $G$ are negligibly small
when the lead polarization varies in the region $p<1$. This observation is, as
a matter of fact, the same as in the QD \cite{GangSu}, where the resonant
probability is independent of the lead-polarization $p$ \cite{JianWang}. There
exist five evident resonant-peaks in Fig. 3(b), in which the first
(corresponding to the transition $\left\vert 0,2;\pm2\right\rangle
\Leftrightarrow\left\vert 1,5/2;\pm5/2\right\rangle $) and fifth ( $\left\vert
1,5/2;\pm5/2\right\rangle \Leftrightarrow\left\vert 2,2;\pm2\right\rangle $)
peaks are higher than the other three resulted from the transitions:
$\left\vert 0,2;\pm1\right\rangle \Leftrightarrow\left\vert 1,3/2;\pm
3/2\right\rangle $ (2) , $\left\vert 0,2;\pm2\right\rangle \Leftrightarrow
\left\vert 1,3/2;\pm3/2\right\rangle $ (3) and $\left\vert 1,3/2;\pm
3/2\right\rangle \Leftrightarrow\left\vert 2,2;\pm2\right\rangle $ (4),
respectively\cite{JB3}. However, for the full polarization $p=1$, the current
and differential conductance are quite different\textbf{ }since only one
spin-up channel is involved in the transition without competition between
majority and minority spin channels, which leads to dynamic spin-blockade
\cite{Timm3,Belzig}. The bias-voltage dependence of the spin current $I_{s}$
and current-polarization $\chi$ for various lead-polarizations $p$ are shown
in Fig. 3(c) and Fig. 3(d). The spin current exhibits step-like behavior in
parallel lead-configurations, which is\ also the same as in a single QD
\cite{GangSu}, while with more steps due to the complex energy levels of SMM.
In addition, when lead-polarization $p$ increases, the spin current varies
more evidently along with an increasing magnitude of current polarization
$\chi$ in contrast to the charge current. The current polarization $\chi$
approaches the largest value at a very low bias voltage, which is dominated by
elastic co-tunneling processes through majority-majority and minority-minority
spin channels. With increasing bias voltage, the inelastic co-tunneling starts
to enter the transport and leads to the decrease of $\chi$. Moreover, as the
bias voltage reaches the threshold of sequential transport, the value of
$\chi$ approaches the lead-polarization $p$ \cite{Jauho2}, due to the
appearance of sequential transport channels.

The results of antiferromagnetic exchange interaction (Fig.4) are different
from the ferromagnetic case due to the two degenerate ground-states
$\left\vert 1,3/2;\pm3/2\right\rangle $ seen from Fig. 2(b), where the
differential conductance spectra have four evident peaks. With the increase of
lead-polarization $p$ (except $p=1$) the position of peak$-2$ shifts toward
the negative direction of the $V$-axis since the contribution is mainly from
the transition of $\left\vert 0,2;\pm1\right\rangle \Leftrightarrow\left\vert
1,5/2;\pm3/2\right\rangle $ instead of $\left\vert 0,2;\pm2\right\rangle
\Leftrightarrow\left\vert 1,5/2;\pm5/2\right\rangle $. The other three peaks
(peak$-1,-3,-4$) are resulted from the transitions of $\left\vert
0,2;\pm2\right\rangle \Leftrightarrow\left\vert 1,3/2;\pm3/2\right\rangle $,
$\left\vert 1,5/2;\pm5/2\right\rangle \Leftrightarrow\left\vert 2,2;\pm
2\right\rangle $, and $\left\vert 1,3/2;\pm3/2\right\rangle \Leftrightarrow
\left\vert 2,2;\pm2\right\rangle $, respectively. For the case of $p=1$, the
position of peak-2 has a great change different from the peak-1 in the
ferromagnetic coupling, where the state $\left\vert 0,2;2\right\rangle $ does
not participate in transition. Besides, $I_{s}$ and $\chi$ exhibit similar
behaviors with the ferromagnetic case.

\subsection{Antiparallel configuration and current polarization reversal}

\begin{figure}[ptb]
\centering \vspace{0cm} \hspace{0cm}
\scalebox{1.1}{\includegraphics{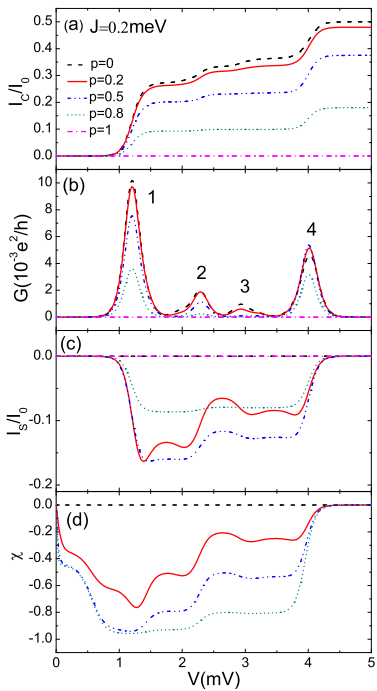}}\caption{(Color online) (a) Charge
current $I_{c}$, (b) differential conductance $G$, (c) spin current $I_{s}$,
and (d) current polarization $\chi$ as a function of the bias voltage $V$ for
different lead-polarizations $p$ with ferromagnetic exchange coupling
($J=0.2meV$) and antiparallel lead-configuration.}%
\label{fig5}%
\end{figure}

In the antiparallel configuration of magnetic leads peculiar behavior of the
spin current is observed, which is shown in Fig. 5. From Fig. 5(a), it is seen
that $I_{c}$ decreases monotonously from a maximum value to zero at a certain
bias-voltage when $p$ increases from $0$ to $1$. Since transport occurs mainly
through two spin channels (majority-minority and minority-majority), for a
large $p$ the most probable transport process is that a spin-down electron of
higher rate from the lead $R$ \ tunnels through the molecule along with a
spin-flip and molecular spin-state lowering simultaneously. However, this
process is limited by the lowest spin-state of molecule and thus the charge
current finally vanishes \cite{Timm3}. Also, the $G$ peaks in Fig. 5(b) are
suppressed with the increase of $p$ and the peak-2 appeared in Fig. 3(b) even
disappears because the transition channels through states $\left\vert
0,2;\pm1\right\rangle $ are blocked.

\begin{figure}[ptb]
\centering \vspace{0cm} \hspace{0cm}
\scalebox{1.1}{\includegraphics{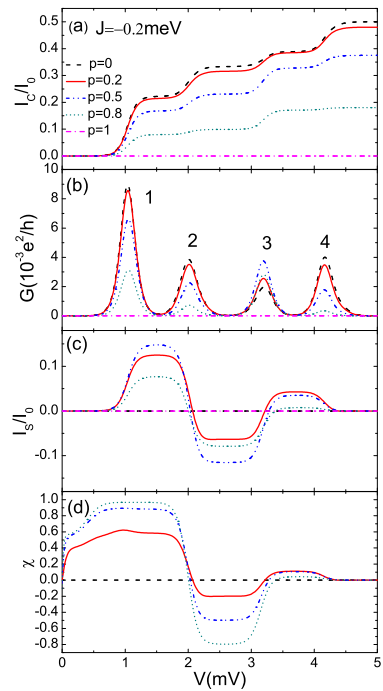}}\caption{((Color online) (a) Charge
current $I_{c}$, (b) differential conductance $G$, (c) spin current $I_{s}$,
and (d) current polarization $\chi$ as a function of the bias voltage $V$ for
different lead-polarizations $p$ with antiferromagnetic exchange coupling
($J=-0.2meV$) and antiparallel lead-configuration.}%
\label{fig6}%
\end{figure}

The spin current $I_{s}$ (see Fig. 5(c)) shows a basin-like behavior in the
region $I_{\uparrow}<I_{\downarrow}$ similar to the QD case. The difference is
that $I_{s}$ in the SMM can be suppressed by the lead-polarization $p$ at some
value, however it is enhanced always with the increase of $p$ ($p\neq1$) in
the QD. Nevertheless, $\chi$ displays a monotonic behavior with increase of
$p$ ($p\neq1$) for a fixed bias voltage $V$ seen from Fig. 5(d). It is found
that the polarization $\chi$ can have a high value at low lead-polarization
$p$ resulted from the exchange coupling between the electron-spin and the SMM.
Moreover, in contrast to the parallel lead-configuration the current with
$\chi=0$ is contributed mainly from elastic co-tunneling processes around zero
bias voltage, which belong to the majority-minority and minority-majority spin
channels. With increasing $V$, the inelastic co-tunneling dominates the
transport and $\chi$ decreases. Beyond the threshold value of sequential
transport the peak-1 arises and $\chi$ reaches the lowest value due to the
transition $\left\vert 0,2;-2\right\rangle \Leftrightarrow\left\vert
1,5/2;-5/2\right\rangle $. Then $\chi$ increases since more transport channels
enter the bias window.

Fig. 6 is the $I_{c}-V$ diagram for antiferromagnetic exchange coupling, in
which the $I_{c}$ curve (Fig. 6(a)) exhibits a similar behavior as in the
ferromagnetic coupling case. The corresponding differential conductance $G$ is
shown in Fig. 6(b). Moreover, in the antiparallel lead-configuration the
steady state probabilities with the negative eigenvalues of spin operator
$S_{tot}^{z}$ become larger than that of positive eigenvalues resulted by the
spin-flip process\cite{Timm3,Martin2,JB3}\textbf{. }Therefore, the peak-1 and
peak-3 of the $G$ curve are mainly contributed by the spin-up electron
transitions $\left\vert 0,2;-2\right\rangle \Leftrightarrow\left\vert
1,3/2;-3/2\right\rangle $ and $\left\vert 1,5/2;-5/2\right\rangle
\Leftrightarrow\left\vert 2,2;-2\right\rangle $ respectively.\textbf{ }On the
other hand, the peak-2 (peak-4) is mainly induced by the transition
$\left\vert 0,2;-2\right\rangle \Leftrightarrow\left\vert
1,5/2;-5/2\right\rangle $ ($\left\vert 1,3/2;-3/2\right\rangle \Leftrightarrow
\left\vert 2,2;-2\right\rangle $) of the spin-down electron. The spin-current
curve $I_{s}$ in Fig. 6(c) has a very interesting characteristic different
from the ferromagnetic case that it possesses an alternate structure of
two-plateau and one-basin\textbf{. }The first plateau appears between the
positions of peak-1 and peak-2, where the spin-up current is larger than the
spin-down (positive $\chi$).\textbf{ }In addition, the spin current shows a
non-monotonic behavior with increasing lead-polarizations $p$, which is
resulted from the competition between the spin-up and spin-down
currents.\textbf{ }Furthermore, with increasing the bias voltage the basin
emerges when the transition $\left\vert 0,2;-2\right\rangle \Leftrightarrow
\left\vert 1,5/2;-5/2\right\rangle $ starts to participate in the transport,
in which the spin-down current becomes larger than the spin-up current
(negative $\chi$).\textbf{ }The second plateau of spin current appears as the
bias voltage increases to the value between the peak-3 and peak-4. Finally
when all transport channels open up, the spin current vanishes. In conclusion,
the polarization of spin current $\chi$ can be inverted by adjusting the bias
voltage and the spin current exhibits rich multi-NDC behaviors \cite{Hua}. The
polarization reversal of spin current may have potential application in the
spintronic device.

\section{Conclusion}

In summary, the spin current through a SMM weakly coupled to two ferromagnetic
leads is obtained in the nonlinear tunneling regime for both ferromagnetic and
antiferromagnetic exchange-couplings. The complex energy spectrum of the SMM
results in rich properties of the currents and conductances compared with the
single QD, which on the other hand provide magnetic signatures of the SMM. The
most interesting observation is the spin-polarization reversal in the
antiferromagnetic exchange-coupling case, which can be manipulated by the bias
voltage. These theoretical results may be useful in the future development of
spintronic devices.

\section{Acknowledgment}

This work was supported by National Natural Science Foundation of China (Grant
No. 11075099, No. 11004124 and No. 10974124).


\begin{thebibliography}{99}                                                                                               %


\bibitem {exp1}H. B. Heersche \textit{et al.}, Phys. Rev. Lett. 96, 206801 (2006).

\bibitem {exp2}M.-H. Jo \textit{et al.}, Nano Lett. 6, 2014 (2006).

\bibitem {exp4}A. S. Zyazin \textit{et al.}, Nano Lett. 10, 3307 (2010).

\bibitem {exp3}J. E. Grose \textit{et al.}, Nat. Mater. 7, 884 (2008).

\bibitem {Timm0}N. Roch \textit{et al.}, Phys. Rev. B 83, 081407(R) (2011).

\bibitem {Kim}G.-H. Kim and T.-S. Kim, Phys. Rev. Lett. 92, 137203 (2004); C.
Romeike, M. R. Wegewijs, and H. Schoeller, Phys. Rev. Lett. 96, 196805 (2006);
J. Fern\'{a}ndez-Rossier and R. Aguado, Phys. Rev. Lett. 98, 106805 (2007); J.
Lehmann and D. Loss, Phys. Rev. Lett. 98, 117203 (2007); C. Romeike \textit{et
al.}, Phys. Rev. B 75, 064404 (2007).

\bibitem {Timm1}F. Elste and C. Timm, Phys. Rev. B 71, 155403 (2005).

\bibitem {Timm2}C. Timm and F. Elste, Phys. Rev. B 73, 235304 (2006).

\bibitem {Timm3}F. Elste and C. Timm, Phys. Rev. B 73, 235305 (2006).

\bibitem {JB2}M. Misiorny and J. Barna\'{s}, Europhys. Lett. 89, 18003 (2010).

\bibitem {Timm4}F. Elste and C. Timm, Phys. Rev. B 75, 195341 (2007).

\bibitem {Kondo}C. Romeike \textit{et al.}, Phys. Rev. Lett. 96, 196601
(2006); C. Romeike \textit{et al.}, Phys. Rev. Lett. 97, 206601 (2006); M. N.
Leuenberger and E. R. Mucciolo, Phys. Rev. Lett. 97, 126601 (2006); G.
Gonzalez, M. N. Leuenberger, and E. R. Mucciolo, Phys. Rev. B 78, 054445
(2008); R.-Q. Wang and D. Y. Xing, Phys. Rev. B 79, 193406 (2009); F. Elste
and C. Timm, Phys. Rev. B 81, 024421 (2010).

\bibitem {JB4}M. Misiorny, I. Weymann, and J. Barna\'{s}, Phys. Rev. B 84,
035445 (2011); M. Misiorny, I. Weymann, and J. Barna\'{s}, Phys. Rev. Lett.
106, 126602 (2011).

\bibitem {Leuenberger}G. Gonz\'{a}lez and M. N. Leuenberger, Phys. Rev. Lett.
98, 256804 (2007).

\bibitem {Martin1}K.-I. Imura, Y. Utsumi, and T. Martin, Phys. Rev. B 75,
205341 (2007); H.-B. Xue, Y.-H. Nie, Z.-J. Li, and J.-Q. Liang, J. Appl. Phys.
108, 033707 (2010); H.-B. Xue, Y.-H. Nie, Z.-J. Li, and J.-Q. Liang, J. Appl.
Phys. 109, 083706 (2011); H.-B. Xue, Y.-H. Nie, Z.-J. Li, and J.-Q. Liang,
Phys. Lett. A 375, 716 (2011).

\bibitem {Martin2}T. Jonckheere, K.-I. Imura, and T. Martin, Phys. Rev. B 78,
045316 (2008).

\bibitem {JB3}M. Misiorny and J. Barna\'{s}, Phys. Rev. B 79, 224420 (2009).

\bibitem {JB11}M. Misiorny and J. Barna\'{s}, Phys. Rev. B 75, 134425 (2007).

\bibitem {JB12}M. Misiorny and J. Barna\'{s}, Phys. Rev. B 76, 054448 (2007).

\bibitem {JB13}M. Misiorny and J. Barna\'{s}, Phys. Rev. B 77, 172414 (2008).

\bibitem {Shen}H.-Z. Lu, B. Zhou, and S.-Q. Shen, Phys. Rev. B 79, 174419 (2009).

\bibitem {Xing}Z. Zhang, L. Jiang, R. Wang, B. Wang, and D. Y. Xing, Appl.
Phys. Lett. 99, 133110 (2011).

\bibitem {RWang}R.-Q. Wang \textit{et al.}, Phys. Rev. Lett. 105, 057202
(2010); Z. Zhang \textit{et al.}, Appl. Phys. Lett. 97, 242101 (2010).

\bibitem {Konig}J. K\"{o}nig and J. Martinek, Phys. Rev. Lett. 90, 166602
(2003); M. Braun, J. K\"{o}nig and J. Martinek, Phys. Rev. B 70, 195345 (2004).

\bibitem {Jauho1}F. M. Souza, A. P. Jauho, and J. C. Egues, Phys. Rev. B 78,
155303 (2008).

\bibitem {Jauho2}F. M. Souza, J. C. Egues, and A. P. Jauho, Phys. Rev. B 75,
165303 (2007).

\bibitem {GangSu}H.-F. Mu, G. Su, and Q.-R. Zheng, Phys. Rev. B 73, 054414 (2006).

\bibitem {Yuan}R. Y. Yuan, R. Z. Wang, and H. Yan, J. Phys. Condens. Matter
19, 376215 (2007).

\bibitem {Bruus}H. Bruus and K. Flensberg, \textit{Many-body Quantum Theory in
Condensed Matter Physics} (Oxford University Press, Oxford, 2004).

\bibitem {Timm5}C. Timm, Phys. Rev. B 77, 195416 (2008).

\bibitem {Koch1}J. Koch, F. von Oppen, Y. Oreg, and E. Sela, Phys. Rev. B 70,
195107 (2004).

\bibitem {Koch2}J. Koch, F. von Oppen, and A. V. Andreev, Phys. Rev. B 74,
205438 (2006).

\bibitem {Matveev}M. Turek and K. A. Matveev, Phys. Rev. B 65, 115332 (2002).

\bibitem {Wegewijs}S. Koller, M. Grifoni, M. Leijnse, and M. R. Wegewijs,
Phys. Rev. B 82, 235307 (2010).

\bibitem {Timm6}C. Timm, Phys. Rev. B 83, 115416 (2011).

\bibitem {Weymann}I. Weymann and J. Barna\'{s}, Eur. Phys. J. B 46, 289 (2005).

\bibitem {JianWang}N. Sergueev, Q. F. Sun, H. Guo, B. G. Wang, and J. Wang,
Phys. Rev. B 65, 165303 (2002).

\bibitem {Belzig}A. Cottet and W. Belzig, Europhys. Lett. 66, 405 (2004); A.
Cottet, W. Belzig, and C. Bruder, Phys. Rev. Lett. 92, 206801 (2004); A.
Cottet, W. Belzig, and C. Bruder, Phys. Rev. B 70, 115315 (2004).

\bibitem {Hua}H.-H. Fu and K.-L. Yao, J. Appl. Phys. 110, 094502 (2011).
\end{thebibliography}
\end{document}